\documentstyle[12pt]{article}
\begin{document}
\newcommand{\ket}[1]{\mbox{$| #1 >$}}
\newcommand{\bra}[1]{\mbox{$< #1 |$}}
\pagestyle{empty}
\newpage

\begin{center}
{\Large \bf  Fonctions de structure polaris\'ees\footnote{\sl Cours donn\'e
\`a la $27^{eme}$ Ecole d'Et\'e de Physique des Particules,
Clermont-Ferrand 1995.} }

ou

{\Large \bf \it  o\`u se cache le spin du proton?}
\vskip 2cm

{\Large \bf  Bernard PIRE }

{\it Centre de Physique Th\'eorique{\footnote {Unit\'e propre 14 du Centre
National
de la Recherche Scientifique.}}, Ecole Polytechnique}

{\it 91128 PALAISEAU Cedex}
\vskip 2cm

\end{center}

Il n'est pas question ici
 de faire une revue extensive de tout ce que l'on
connait des r\'eactions inclusives polaris\'ees mais seulement une br\`eve
pr\'esentation si possible p\'edagogique des fonctions de structure
polaris\'ees
puis quelques remarques sur le sujet neuf du spin transverse[AEL].

\section{D\'efinitions}
\subsection{Factorisation leptonique-hadronique }

On consid\`ere la diffusion in\'elastique de  leptons polaris\'es
longitudinalement
sur des nucl\'eons polaris\'es.
On note $m$ la masse du lepton ,  $ k~(k^\prime$) la 4-impulsion du lepton
initial (final)
et $s~(s^\prime)$ son 4-vecteur spin covariant , tel que
 $s \cdot  k$ = 0 $(s^\prime \cdot  k^\prime = 0)$ et $s \cdot s =
- 1$ $(s^\prime \cdot s^\prime = -1)$; la masse du nucl\'eon est $M$,
sa 4-impulsion et son spin sont
 $P$ et $S$. On suppose la dominance de l'\'echange d'unseul photon
et la section efficace
differentielle pour d\'etecter le lepton final polaris\'e
dans l'angle solide $d\Omega$ et dans l'intervalle d'\'energie
 $(E^\prime,~E^\prime +
dE^\prime)$ dans le rep\`ere du laboratoire, $P = (M, 0), ~k = (E,\vec{k}),
{}~k^\prime = (E^\prime,\vec{{k}^\prime})$, s'\'ecrit comme:
\begin{equation}
{d^2\sigma\over d\Omega~dE^\prime} = {\alpha^2\over 2 M q^4}~{E^\prime\over
E}~L_{\mu\nu}~W^{\mu\nu} \,,
\end{equation}
avec $q =  k- k^\prime$ .

Dans l'Eq. (1) le  tenseur leptonique $L_{\mu\nu}$ est
\begin{equation}
L_{\mu\nu} ( k, s; k^\prime, s^\prime) = [\bar{u} (k^\prime, s^\prime)
{}~\gamma_{\mu}~u(k,s)]^\ast
{}~[\bar{u}(k^\prime, s^\prime)~\gamma_{\nu}~u(k,s)]
\end{equation}
que l'on s\'epare en parties sym\'etrique $(S)$ et antisym\'etrique $(A)$
 par rapport \`a l'interchange
$\mu,\nu$ :
\begin{eqnarray}
L_{\mu\nu}(k, s; k^\prime, s^\prime)&=& L^{(S)}_{\mu\nu}~(k;k^\prime) +
iL^{(A)}_{\mu\nu}~(k,s;k^\prime) \\
&+&L^{\prime~(S)}_{\mu\nu}~(k,s;k^\prime,s^\prime) +
iL^{\prime~(A)}_{\mu\nu}~(k;k^\prime, s^\prime)\nonumber
\end{eqnarray}
avec
\begin{eqnarray}
L^{(S)}_{\mu\nu}~(k;k^\prime) &=& k_\mu k^\prime_\nu + k^\prime_\mu k_\nu -
g_{\mu\nu}~(k\cdot k^\prime - m^2)\\
L^{(A)}_{\mu\nu}~(k,s;k^\prime) &=&
m~\varepsilon_{\mu\nu\alpha\beta}~s^\alpha~(k-k^\prime)^{\beta}\\
L^{\prime~(S)}_{\mu\nu}(k,s;k^\prime,s^\prime)&=&(k\cdot s^\prime)~
(k^\prime_\mu s_\nu +
s_{\mu}k^\prime_\nu-g_{\mu\nu}~k^\prime\cdot s) \nonumber\\
&-&(k\cdot k^\prime - m^2)~(s_\mu s^\prime_\nu + s^\prime_\mu s_\nu -
g_{\mu\nu}~s\cdot s^\prime) \\
&+&(k^\prime \cdot s)(s^\prime_\mu k_\nu + k_\mu s^\prime_\nu) -
(s \cdot s^\prime)(k_\mu k^\prime_\nu + k^\prime_\mu k_\nu)\nonumber\\
L^{\prime~(A)}_{\mu\nu}~(k;k^\prime,s^\prime) &=&
m~\varepsilon_{\mu\nu\alpha\beta}
{}~s^{\prime\alpha} (k-k^\prime)^{\beta} \,.
\end{eqnarray}

Si on somme l'Eq. (3) par rapport \`a $s^\prime$ et qu'on moyenne par
rapport \`a $s$ on retrouve
le tenseur leptonique non polaris\'e , 2$L^{(S)}_{\mu\nu}$. Si on ne somme
que sur
$s^\prime$, on obtient $2L^{(S)}_{\mu\nu} + 2iL^{(A)}_{\mu\nu}$.

Le tenseur hadronique $W_{\mu\nu}$ est  d\'efini de la m\^eme fa\c con en
termes
de quatre fonctions de structure :
\begin{equation}
W_{\mu\nu}(q; P,S) = W^{(S)}_{\mu\nu}(q;P) + i~W_{\mu\nu}^{(A)}(q; P, S)
\end{equation}
avec
\begin{eqnarray}
{1\over 2M}~W^{(S)}_{\mu\nu}(q;P) &=& \left(-g_{\mu\nu} + {q_\mu q_\nu\over
q^2}\right)~W_1(P\cdot q, q^2) \nonumber\\
&+&\left[\left(P_\mu -{P\cdot q\over q^2}~q_\mu\right)\left(P_\nu -
{P\cdot q\over q^2}~q_\nu\right)\right]{W_2(P\cdot q, q^2)\over M^2}\\
{1\over 2M}~W^{(A)}_{\mu\nu} (q;P,S) &=&
\varepsilon_{\mu\nu\alpha\beta}~q^\alpha\Biggl\{M S^\beta G_1 (P\cdot q, q^2)
\nonumber\\
&+&[(P\cdot q) S^\beta - (S\cdot q) P^\beta]~{G_2 (P \cdot q, q^2)\over M}
\Biggl\} \,.
\end{eqnarray}

{}On a donc
\begin{eqnarray}
{d^2\sigma\over d\Omega~dE^\prime}&=&{\alpha^2\over 2M q^4}~{E^\prime\over E}~
\left[ L^{(S)}_{\mu\nu}~W^{\mu\nu(S)} +
L^{\prime~(S)}_{\mu\nu}~W^{\mu\nu (S)} \right.\nonumber\\
&-&\left. L^{(A)}_{\mu\nu}~W^{\mu\nu(A)} - L^{\prime~(A)}_{\mu\nu}~
W^{\mu\nu(A)}\right] \,.
\end{eqnarray}

On peut \'etudier chaque terme en consid\'erant des sections efficaces
ou des diff\'erences de sections efficaces.
Par exemple, la section efficace non polaris\'ee habituelle est
proportionnelle \`a $L^{(S)}_{\mu\nu}~W^{\mu\nu(S)}$
\begin{eqnarray}
{d^2\sigma^{unp}\over d\Omega~dE^\prime}\left(k, P;k^\prime\right) &=&{1\over
4}\sum_{s,s^\prime,S}~{d^2\sigma\over d\Omega~d E^\prime}~(k, s, P, S;
k^\prime, s^\prime) \nonumber\\
&=&{\alpha^2\over 2M q^4}~{E^\prime\over E}~2L_{\mu\nu}^{(S)}~W^{\mu\nu(S)}\,,
\end{eqnarray}
tandis que des diff\'erences de sections efficaces avec des valeurs
 oppos\'ees de  spins de la cible ne d\'ependent que du terme
$L_{\mu\nu}^{(A)}~W^{\mu\nu(A)}$ :
$$
\sum_{s^\prime}\left[{d^2\sigma\over d\Omega~dE^\prime}~(k, s, P, - S;
k^\prime, s^\prime)
-{d^2\sigma\over d\Omega ~dE^\prime}~(k, s, P, S; k^\prime, s^\prime)\right]
$$
\begin{equation}
={\alpha^2\over 2Mq^4}~{E^\prime\over E}~4L_{\mu\nu}^{(A)}~W^{\mu\nu(A)}\,.
\end{equation}

\subsection{Fonctions de Structure et scaling de  Bjorken}
\vskip 6pt

La section efficace non polaris\'ee s'\'ecrit
\begin{equation}
{d^2\sigma^{unp}\over d\Omega~dE^\prime} = {4\alpha^2 E^{\prime 2}\over q^4}
\Biggl[2W_1 \sin^2 {\theta\over 2} + W_2 \cos^2{\theta\over 2}\Biggl]
\end{equation}
o\`u $\theta$ est l'angle de diffusion (dans le laboratoire) du lepton.
Cela permet de mesurer les fonctions de structure:
$W_1(P\cdot q, q^2)$ and $W_2 (P\cdot q, q^2).$

Dans la limite de Bjorken ( Deep Inelastic Scattering (DIS) regime),

\begin{equation}
-q^2 = Q^2 \to \infty\quad\quad
\nu = E - E^\prime \to \infty \quad\quad x = {Q^2\over 2 P\cdot q} =
{Q^2\over 2M\nu} \,, \>{\mbox{\rm fix\'e}}
\end{equation}
on a une invariance d'\'echelle:
\begin{eqnarray}
\lim_{Bj}~MW_1(P\cdot q, Q^2) &=& F_1 (x)\nonumber\\
\lim_{Bj}~\nu W_2(P\cdot q, Q^2) &=& F_2 (x) \,,
\end{eqnarray}
o\`u $F_{1,2}$ varie tr\`es lentement avec $Q^2$ \`a  $x$ fix\'e.

De la m\^eme fa\c con, on a
$$\sum_{s^\prime}\left[{d^2\sigma\over d\Omega~dE^\prime}(k, s, P, S; k^\prime,
s^\prime) - {d^2\sigma\over d\Omega~d E^\prime} (k, s, P - S; k^\prime,
s^\prime) \right]\equiv$$
\begin{equation}
\equiv {d^2\sigma^{s,S}\over d\Omega~d E^\prime} -
{d^2 \sigma^{s, -S}\over d\Omega~dE^\prime}~~=
\end{equation}
$$={8m \alpha^2 E^\prime\over q^4E}\left\{\left[(q\cdot S) (q\cdot s) +
Q^2 (s\cdot S)\right] MG_1 + Q^2 \Bigl[
(s\cdot S)(P\cdot q)-(q\cdot S)(P\cdot s)\Bigr]{G_2\over M}\right\}$$
qui permet de mesurer les fonctions de structure polaris\'ees
$G_1 (P\cdot q, q^2)$ and $G_2 (P\cdot q, q^2)$. Elles aussi "scale"
approximativement:
\begin{eqnarray}
\lim_{Bj}~{(P\cdot q)^2\over \nu}~G_1(P\cdot q, Q^2)&=& g_1 (x)\\
\lim_{Bj}~\nu ~(P\cdot q)~G_2 (P\cdot q, q^2) &=& g_2 (x) \,. \nonumber
\end{eqnarray}
En termes of $g_{1,2}$ l'expression de $W^{(A)}_{\mu\nu}$ devient
\begin{equation}
W^{(A)}_{\mu\nu }(q; P, s) = {2M\over P\cdot q}~\varepsilon_{\mu v\alpha\beta}~
q^\alpha\Biggl\{ S^\beta g_1 (x, Q^2)+ \left[S^\beta
- {(S\cdot q) \, P^\beta\over (P\cdot q)}\right] g_2 (x, Q^2)\Biggl\}\,.
\end{equation}

\subsection{Comment mesurer $g_1$ ?}
\vskip 6pt

 Lorsque les nucl\'eons sont polaris\'es le long $(\Rightarrow)$ ou oppos\'e
($\Leftarrow$) \`a la direction du  lepton initial, on a
\begin{equation}
{d^2\sigma^{\begin{array}{c}\hspace*{-0.2cm}\to\vspace*{-0.3cm}\\
\hspace*{-0.2cm}\Rightarrow\end{array}}\over d\Omega~dE^\prime} -
{d^2\sigma^{\begin{array}{c}\hspace*{-0.2cm}\to\vspace*{-0.3cm}\\
\hspace*{-0.2cm}\Leftarrow\end{array}}\over
d\Omega~dE^\prime} =
-{4\alpha^2\over Q^2}~{E^\prime\over E}\Biggl[(E + E^\prime\cos\theta)MG_1 -
Q^2G_2\Biggl] \,.
\end{equation}

Si les  nucl\'eons sont polaris\'es de fa\c con transverse ,
le spin du nucl\'eon \'etant perpendiculaire \`a la direction du lepton
entrant, on a:
\begin{equation}
{d^2 \sigma^{\to\Uparrow}\over d\Omega~dE^\prime} -
{d^2 \sigma^{\to\Downarrow}\over d\Omega~dE^\prime} =
- {4 \alpha^2\over Q^2}~{E^{\prime 2}\over E}~\sin\theta \cos\phi ~(MG_1 +
2EG_2)\,.
\end{equation}

On r\'e\'ecrit souvent l'assym\'etrie:
\begin{equation}
{M \nu Q^2E\over 2 \alpha^2E^\prime(E+E^\prime \cos\theta)}~{d^2
\sigma^{unp}\over d
\Omega~dE^\prime}~A_{\parallel} = g_1 - {2xM\over E +
E^\prime \cos\theta}~g_2
\end{equation}
soit
\begin{equation}
g_1 - \kappa g_2 = 2 K~d\sigma^{unp}~A_{\parallel}
\end{equation}
avec
$$\kappa = {2xM\over E + E^\prime \cos\theta} \approx {xM\over E -
Q^2/(4 Mx)}$$
\begin{equation}
K = {M \nu Q^2 E\over 4 \alpha^2 E^\prime (E + E^\prime \cos\theta)} =
{EE^\prime~\cos^2
(\theta/2) \over 2x\sigma_{Mott}~(E + E^\prime \cos\theta)}
\end{equation}
o\`u
$$\sigma_{Mott} = \Biggl[{\alpha \cos(\theta/2)\over 2E \sin^2(\theta/2)}
\Biggr]^2 \cdot$$

La mesure de
 $A_{\parallel}$ (et de
$d\sigma^{unp}$)nous permet donc d'extraire la  combinaison $g_1 - \kappa g_2$.
Il se trouve qu'on peut dans un premier temps n\'egliger le terme en $g_2$
car le coefficient cin\'ematique $\kappa$
est minuscule \`a haute \'energie. On en conclut que la mesure de
l'assym\'etrie
longitudinale est une d\'etermination satisfaisante  de la fonction de
 structure $g_1$.

\setcounter{section}{1}
\section{le Mod\`ele des Partons dans le  DIS polaris\'e}
\vskip 6pt
Dans sa version la plus simple (voir par exemple [LP]), le Mod\`ele des
Partons d\'ecrit le proton
comme une superposition de constituants libres colin\'eaires, transportant
chacun une  fraction $x^\prime$
de la 4-impulsion du nucl\'eon. La diffusion in\'elastique profonde
lepton-nucl\'eon
est alors d\'ecrite comme la somme incoh\'erente d'interactions lepton-quark
et le tenseur hadronique $W_{\mu\nu} (N)$ s'exprime en termes
de tenseurs \'el\'ementaires   $w_{\mu\nu}$ calcul\'es au niveau des quarks:
\begin{eqnarray}
W_{\mu\nu} (q;P,S) &=& W^{(S)}_{\mu\nu} (q;P) + i W^{(A)}_{\mu\nu} (q;P,S)
\nonumber\\
&=&
\sum_{q,s} e^2_q \, \frac{1}{2P\cdot q} \int^1_0 \frac{dx^\prime}{x^\prime}~
\delta (x^\prime-x)~n_q(x^\prime,s;S) ~w_{\mu\nu} (x^\prime,q,s) \,,
\end{eqnarray}
o\`u $n_q (x^\prime,s;S)$ est la densit\'e des quarks $q$ de charge
$e_q$, fraction  de 4-impulsion $x^\prime$ et spin $s$ dans un  nucl\'eon de
spin $S$ et de 4-impulsion $P$; la somme $\Sigma_q$ agit sur les quarks et les
 antiquarks;
$x$ est la variable de Bjorken  et le
 tenseur $w_{\mu\nu} (x,q,s)$
est d\'eduit du  tenseur leptonique $L_{\mu\nu}$ en
rempla\c cant $k^\mu\ \to x P^\mu, ~k^{\prime \mu} \to xP^\mu + q^\mu$ et en
sommant sur les \'etats de spin  $(s^\prime)$ non observ\'es du quark final.
On a donc:
\begin{equation}
w_{\mu\nu} (x,q,s) = w^{(S)}_{\mu\nu} (x,q)  + i w^{(A)}_{\mu\nu} (x,q,s)
\end{equation}
avec
\begin{eqnarray}
w^{(S)}_{\mu\nu} (x,q) &=& 2\,[2x^2 P_\mu P_\nu + xP_\mu q_\nu + xq_\mu P_\nu
- x(P\cdot q) g^{\mu\nu}]\\
w_{\mu\nu}^{(A)} (x,q,s) &=& -2m_q ~ \varepsilon_{\mu\nu\alpha\beta}
{}~s^\alpha q^\beta
\end{eqnarray}
et on doit prendre pour \^etre coh\'erent la masse du quark $m_q = xM$,
 avant et apr\`es l'interaction avec le  photon virtuel.

{} On obtient de ces \'equations les pr\'edictions du Mod\`ele des Partons Naif
pour les fonctions de structure non polaris\'ees du nucl\'eon:
\begin{eqnarray}
F_1 (x) &=& \frac{1}{2} \sum_q e^2_q ~q(x) \\
F_2 (x) &=& x \sum_q e^2_q ~q(x) = 2x F_1 (x) \,,
\end{eqnarray}
o\`u les densit\'es non polaris\'ees de nombre de quarks  $q(x)$ sont
d\'efinies
comme
\begin{equation}
q(x) = \sum_s n_q (x,s;S)\,.
\end{equation}

On obtient de la m\^eme fa\c con les fonctions de structure polaris\'ees:
\begin{eqnarray}
g_1(x) &=& \frac{1}{2} \sum_q e^2_q ~\Delta q(x,S) \\
g_2(x) &=& 0
\end{eqnarray}
avec
\begin{equation}
\Delta q(x,S) = n_q (x,S;S) - n_q (x,-S;S)
\end{equation}
la diff\'erence entre la densit\'e de quarks avec le spin parall\`ele
 au spin du  nucl\'eon  ($s=S$) et celle avec le spin anti-parall\`ele
($s=-S$).

Le fait que $g_2$ soit nul dans le mod\`ele des partons montre qu'il sera
difficile
de se faire une image physique des contributions \`a cette quantit\'e. On
ne reviendra
pas sur ce point d\'elicat mais extr\`emement int\'eressant dans ce cours.

\section{R\`egles de somme }

Si on int\`egre sur $x$ la relation (32), on obtient

\begin{equation}
\int^1_0 g_1(x) dx = \frac{1}{2} \sum_q e^2_q ~\Delta q
\end{equation}

avec

\begin{equation}
\Delta q = \int^1_0 ( n_q (x,S;S) - n_q (x,-S;S) )
\end{equation}
et donc
\begin{equation}
\int^1_0 g^p_1(x) dx = \frac{2}{9}  ~\Delta u + \frac{1}{18}  ~\Delta d +
\frac{1}{18}  ~\Delta s
\end{equation}

\subsection{La r\`egle de somme de Bjorken }

Lorsqu'on fait la diff\'erence entre les int\'egrales de $g_1$ pour le proton
et
le neutron, on obtient:
\begin{equation}
\int^1_0 ( g^p_1(x) - g^n_1(x) ) dx = \frac{1}{6} ( ~\Delta u -   ~\Delta d )
\end{equation}

Il se trouve que les techniques de l'alg\`ebre des courants et les
propri\'et\'es
d'invariance par rapport aux rotations d'isospin des courants
\'electromagn\'etiques
et faibles permettent de relier cette diff\'erence aux param\^etres de la
 d\'esint\'egration $\beta$ du neutron. Bjorken [BJO] a ainsi \'etabli que
\begin{equation}
\int^1_0 ( g^p_1(x) - g^n_1(x) ) dx = \frac{1}{6} \frac{g_A}{g_V}
\end{equation}

Cette conclusion n'est pas alt\'er\'ee lorsqu'on se place dans le cadre de
la QCD,
\`a de petites corrections pr\`es. La v\'erification exp\'erimentale de cette
r\`egle de somme est \'evidemment un d\'efi puisqu'il faut extraire la fonction
de structure du neutron. On utilise pour cela des noyaux de deuterium ou
d'helium3.
Les r\'esultats d'une telle analyse sont montr\'es sur la figure 1.

\vskip10cm

\begin{center}
Fig.1: Test exp\'erimental de la r\`egle de somme de Bjorken.
\end{center}

\subsection{La r\`egle de somme de Ellis-Jaffe }

Il est naturel d'introduire les quantit\'es singlet,triplet et octet (selon
$SU(3)_{saveur}$) d\'efinies dans le mod\`ele des quarks par:

\begin{equation}
a_0 = \Delta\Sigma\equiv \int^1_0 dx~\Delta\Sigma (x)
\end{equation}

\noindent
avec
\begin{equation}
\Delta\Sigma (x)\equiv\Delta u(x)+\Delta\bar{u}(x)+\Delta d(x)+
\Delta\bar{d}(x)+\Delta s(x) + \Delta\bar{s}(x)\,.
\end{equation}

\begin{equation}
a_3=\int^1_0 dx~[\Delta u(x)+\Delta \bar{u}(x)-\Delta d(x)-
\Delta\bar{d} (x)]
\end{equation}

et
\begin{equation}
a_8={1\over \sqrt{3}} \int^1_0 dx~[\Delta u(x)+\Delta\bar{u}(x)+
\Delta d(x)+\Delta\bar{d}(x)-2\Delta s(x)-2\Delta\bar{s}(x)].
\end{equation}

Si on suppose maintenant que l'on peut n\'egliger la contribution de la mer
\'etrange, on obtient la  r\`egle de somme de Ellis-Jaffe [EJ]:

\begin{equation}
\Gamma^p_1 \equiv\int^1_0 g^p_1(x) dx = {1\over 12} \left\{ a_3 +
{5\over \sqrt 3}\, a_8 \right\} \simeq 0.188 \pm 0.004
\end{equation}
o\`u la valeur num\'erique vient de la d\'etermination \`a partir des
d\'esint\'egrations des hyp\'erons des quantit\'es $a_3$ et $a_8$ .
Sans entrer dans les d\'etails, disons que les hypoth\`eses permettant
d'extraire
ces quantit\'es sont principalement:

\begin{description}
\item{a)} que les huit hyp\'erons de spin 1/2  forment un octet selon
$SU(3)_F$;
\item{b)} que les  courants $J^j_\mu,~J^j_{5\mu}\ (j = 1,...,8)$ se
transforment
comme un  octet sous $SU(3)_F$ , $J^j_\mu,~J^j_{5\mu}$ \'etant  conserv\'es;
\item{c)} que les impulsions transf\'er\'ees et les diff\'erences de masse
dans les transitions hadroniques ont des effets n\'egligibleables;
\end{description}

C'est la violente violation exp\'erimentale de cette r\`egle de somme
qui, sous le nom accrocheur de {\it crise du spin} a r\'eveill\'e l'int\'er\^et
de la communaut\'e pour les variables polaris\'ees.

Des r\'esultats exp\'erimentaux sur
$g_1^p$ ont \'et\'e d'abord obtenus  \`a SLAC [ALG] en 1978 avec un faisceau
d'\'electrons.La collaboration SLAC-Yale continua ce travail en 1983
[BAU] . Plus r\'ecemment, la collaboration EMC (European Muon Collaboration)
[ASH]
a utilis\'e un faisceau de muons
longitudinalement polaris\'es d'\'energie 100--200 GeV sur une cible
d'hydrog\`ene
 longitudinalement
polaris\'e.
La collaboration SMC au CERN [ADA] et les exp\'eriences  E142 - E143 au
SLAC [ANT],
ont enfin affin\'e de fa\c con remarquable les r\'esultats en incluant, pour la
premi\`ere fois, des donn\'ees sur le  neutron.

La Figure 2 montre l'\'etat actuel des donn\'ees.

\vskip10cm

\begin{center}
Fig.2: La fonction de structure $g_1 (x,Q^2)$ mesur\'ee \`a SLAC et au CERN.
\end{center}
On exprime souvent de mani\`ere un peu diff\'erente les m\^emes donn\'ees
en \'ecrivant
\begin{equation}
a_0={3\over 4}\Biggl\{12\Gamma^p_1-a_3-{1\over\sqrt{3}}\,a_8\Biggl\}\ .
\end{equation}
que l'exp\'erience fixe donc \`a:
\begin{equation}
a_0 = 0.06 \pm 0.12 \pm 0.17 \,.
\end{equation}
Un raisonnement dans le cadre du mod\`ele des quarks aurait impliqu\'e
 $a_0 \simeq 1$. D'o\`u la surprise sinon la crise.

On peut m\^eme d\'ecomposer cette structure en spin du proton en utilisant
les valeurs de $a_3$ et $a_8$ venant des d\'esint\'egrations des hyp\'erons.
On obtient:
\begin{eqnarray}
\Delta u&=& ~~0.79 \pm 0.03 \pm 0.04\nonumber\\
\Delta d&=& -0.47 \pm 0.03 \pm 0.04\\
\Delta s&=& -0.26 \pm 0.06 \pm 0.09\nonumber
\end{eqnarray}

\section{Corrections radiatives}

\subsection{Les \'equations d'\'evolution}
On sait bien que les corrections radiatives de QCD induisent des violations
logarithmiques de l'invariance d'\'echelle de Bjorken. L'expression de
 $\Gamma^p_1$ par exemple est modifi\'ee en
\begin{equation}
\Gamma^p_1(Q^2)={1\over 12}\Biggl\{\Biggl(a_3+{1\over\sqrt 3}\,a_8\Biggl)
E_{NS}(Q^2)+{4\over 3}\,a_0\,E_S(Q^2)\Biggl\}
\end{equation}
o\`u les coefficients  $E_{NS}$ and $E_S$ peuvent \^etre calcul\'es
perturbativement [KOD]; on les connait maintenant jusqu'\`a l'ordre
$\alpha^2_s$ and
$\alpha^3_s$ respectivement [LAR]
\begin{equation}
E_{NS}(Q^2) = 1 - {\alpha_s\over \pi} -
3.25\Biggl({\alpha_s\over
\pi}\Biggl)^2 -
13.85\Biggl({\alpha_s
\over \pi}\Biggl)^3
\end{equation}
\begin{equation}
E_{S}(Q^2) = 1 - 0.040
\Biggl({\alpha_s\over \pi}\Biggl) +
0.07
\Biggl({\alpha_s\over\pi}\Biggl)^2
\end{equation}
avec $\alpha_s = \alpha_s(Q^2)$ et $N_f = 3$. On doit tenir compte de ces
 corrections perturbatives lorsqu'on analyse des donn\'ees prises \`a des
  $Q^2$ diff\'erents.

Au niveau des logarithmes dominants, l'expression des
 $a_j$ en termes de densit\'es de quarks $\Delta q(x)$ est simplement
modifi\'ee en rempla\c cant
\begin{equation}
\Delta q(x) \to \Delta q(x;Q^2)
\end{equation}
o\`u  l'\'evolution en $Q^2$ des quantit\'es $\Delta q(x;Q^2)$ est
  contr\^ol\'ee par les \'equations
d'Altarelli-Parisi  [AP] dans leur version spin-d\'ependante.
 Dans le secteur singulet par exemple, elles s'\'ecrivent:
\begin{eqnarray}
{d\Delta\Sigma(x,t)\over d\ln Q^2}
&=\frac{\alpha_s}{2\pi}\int_x^1\! \frac{dy}{y}\,\left[
P_{qq}^{\rm S}(\frac{x}{y},\alpha_s)\Delta\Sigma(y,t)
          + 2n_fP_{qg}(\frac{x}{y},\alpha_s)\Delta g(y,t)\right]\\
{d\Delta g(x,t) \over d\ln Q^2}
&=~~~~\frac{\alpha_s}{2\pi}\int_x^1\! \frac{dy}{y}\,\left[
P_{gq}(\frac{x}{y},\alpha_s)\Delta\Sigma(y,t)
                ~+~ P_{gg}(\frac{x}{y},\alpha_s)\Delta g(y,t)\right]%}xxx
\end{eqnarray}
\noindent
o\`u on a not\'e $P$ ce qui est parfois not\'e $\Delta P$.
 La fonction $\Delta P_{qq}$ est en fait
\'egale \`a  $ P_{qq}$ puisque l'h\'elicit\'e d'un quark
non massif est conserv\'ee lors de l'\'emission d'un gluon.
A ce niveau, la prise en compte des corrections radiatives
n'est rien de plus qu'une modification mineure et on trouve par exemple pour
 $Q^2 \simeq 10$ (GeV/c$)^2$:

\begin{equation}
a_0 = 0.17 \pm 0.12 \pm 0.17
\end{equation}
\noindent
ou, en d\'ecomposant comme plus haut sur les diff\'erentes saveurs:

\begin{eqnarray}
\Delta u&=& ~~0.82 \pm 0.03 \pm 0.04\nonumber\\
\Delta d&=& -0.44 \pm 0.03 \pm 0.04\\
\Delta s&=& -0.21 \pm 0.06 \pm 0.09\nonumber
\end{eqnarray}

Mais il existe un effet plus subtil et en m\^eme temps plus violent des
corrections radiatives: c'est celui li\'e \`a l'anomalie axiale.

\subsection{l'effet de l'anomalie}

La mesure de $\Gamma^p_1$ est interpr\'et\'ee comme une mesure effective de
$a_0$
qui est proportionelle \`a la valeur moyenne dans le proton
du courant axial singulet de saveur $J^0_{5\mu}$

\begin{equation}
J^0_{5\mu} = {\bar{\psi}} \gamma_\mu \gamma_5 {\psi}\,.
\end{equation}

Il se trouve que ce courant, s'il est conserv\'e au niveau classique (en
n\'egligeant les masses des quarks) a une divergence non nulle au niveau
quantique: c'est ce qu'on appelle l'anomalie triangulaire (on n'a
pas la place ici de d\'evelopper la th\'eorie assez subtile de
l'anomalie, voir tout bon livre de th\'eorie des champs) qui s'\'ecrit comme:

\begin{equation}
\partial^\mu J^0_{5\mu} = {{\alpha_s} \over {\pi}} N_f tr F_{\mu \nu}
{\tilde F^{\mu \nu}}.
\end{equation}

L'effet de cette anomalie est de m\'elanger les gluons au courant axial
singulet
des quarks. Par contre, elle ne modifie en rien les courants non singulets.

Il s'ensuit que la relation entre $a_0$ and $\Delta\Sigma$ est
assez diff\'erente de ce que nous disait le mod\`ele naif des partons,
 puisqu'on a maintenant:

\begin{equation}
a_0(Q^2) = \Delta \Sigma - 3\,{\alpha_s(Q^2)\over 2 \pi}\,\Delta g(Q^2)
\end{equation}

On peut introduire  un   {\it courant axial gluonique}
\begin{eqnarray}
K^\mu&=&{1\over 2}\,\varepsilon^{\mu\nu\rho\sigma} A^a_\nu \Biggl(
G^a_{\rho\sigma}-{g\over 3}\,f_{abc} A^b_\rho A^c_\sigma \Biggl)\nonumber\\
&=&\varepsilon^{\mu \nu\rho\sigma} \mbox{\rm Tr} \, \Biggl\{ A_\nu
\Bigg(G_{\rho\sigma} + {i\over 3}\, g[A_\rho, A_\sigma]\Biggl)
\Biggl\}
\end{eqnarray}
avec la matrice $A_\rho={\lambda_a\over 2} A^a_\rho$,
et on a alors
\begin{equation}
\partial_\mu K^\mu = {1\over 2}\,G^a_{\mu\nu}\widetilde{G}^{\mu \nu}_a =
\mbox{\rm Tr}\,(G_{\mu\nu}~\widetilde{G}^{\mu\nu})\,.
\end{equation}
\noindent
Le courant modifi\'e
\begin{equation}
\widetilde{J}^f_{5\mu} \equiv J_{5\mu}^f - {\alpha_s\over 2\pi}~K_\mu
\end{equation}
est donc conserv\'e, $\partial^\mu \widetilde{J}^f_{5\mu} = 0$.

Les \'el\'ements de matrice du courant singlet axial  modifi\'e
\begin{equation}
\widetilde{J}^0_{5\mu} \equiv J^0_{5\mu} - N_f\,{\alpha_s\over 2\pi}\,K_\mu
\end{equation}
devraient correspondre aux valeurs obtenues dans le Mod\`ele des  Quarks
\begin{equation}
\langle P,S|\widetilde{J}_{5\mu}^0|P,S \rangle = 2 M \tilde{a}_0 S^\mu \,,
\end{equation}

Il est important de noter que  $\tilde{a}_0$ est ind\'ependant de $Q^2$.
 Cela provient\footnote { Consid\`erons la  charge axiale $\widetilde{Q}_5$
associ\'ee
 au courant conserv\'e
 $\widetilde{J}^0_{5\mu}$, soit
$\widetilde{Q}_5 = \int d^3 x \, \widetilde{J}^0_{50}(x,t)\,$.
Comment $\widetilde{Q}_5$ peut-il d\'ependre de l'\'echelle de
renormalisation $\mu^2$, dans une th\'eorie sans masse? La seule fa\c con
serait via la variable $\mu t$ et une telle d\'ependance induirait une
 d\'ependance en $t$. Mais on sait que la charge associ\'ee \`a un
 courant local conserv\'e  est ind\'ependante du temps.
 Donc $\widetilde{Q}_5$ doit \^etre ind\'ependant  de l'\'echelle de
renormalisation, et
$\widetilde{a}_0$ est ind\'epenaent de $Q^2$.
Donc  $\Delta\Sigma$ est ind\'ependant de $Q^2$ puisque
$\widetilde{a}_0 = (\Delta\Sigma -3\,{\alpha_s\over 2\pi}\,\Delta g)+
3\,{\alpha_s\over 2\pi}\,\Delta g = \Delta \Sigma$.
}
 de ce que
$\widetilde{a}_0$ est reli\'e au courant conserv\'e
$\widetilde{J}^0_{5\mu}$ et est donc ind\'ependant de l'\'echelle de
 renormalisation $\mu^2$
(ou $Q^2$).

Il est tout \`a fait surprenant que le terme gluonique survive \`a grand
 $Q^2$, puisque cela semble \^etre une correction en $\alpha_s$ qui devrait
dispara\^\i tre lorsque $Q^2\to\infty$. En fait, la fraction de spin emport\'ee
par le  gluon se comporte plut\^ot comme $[\alpha_s(Q^2)]^{-1}$ lorsque
$Q^2\to\infty$. On peut le voir sur les \'equations d'Altarelli-Parisi pour
les premiers moments qui s'\'ecrivent

\begin{equation}
{d \over d\ln Q^2}\Delta\Sigma(Q^2)
=\frac{\alpha_s}{2\pi}\left[
P_{qq}^{\rm S}\Delta\Sigma(Q^2)
          + 2n_fP_{qg}\Delta g(Q^2)\right] \nonumber
\end{equation}
\begin{equation}
{d \over d\ln Q^2}\Delta g(Q^2)
=\frac{\alpha_s}{2\pi}\left[
P_{gq}\Delta\Sigma(Q^2)
                 + P_{gg}\Delta g(Q^2)\right],
\end{equation}
o\`u les deux points cruciaux sont que:

1.$P_{qq}^{\rm S}$ et $P_{qg}$ sont nuls \`a l'ordre d'une boucle.

2.$\Delta\Sigma-{3\alpha_s(Q^2)\over 2\pi}\,\Delta g(Q^2)$ est un vecteur
propre.

\noindent
 On peut donc \'ecrire:
\begin{equation}
{d\over d\ln Q^2}\Biggl[\Delta\Sigma-{3\alpha_s(Q^2)\over 2\pi}\,\Delta g(Q^2)
\Biggl]=-\gamma(\alpha_s)\Biggl[\Delta\Sigma-{3\alpha_s(Q^2)\over 2\pi}\,
\Delta g(Q^2)\Biggl]
\end{equation}
o\`u $\gamma(\alpha_s)$ est la dimension anormale de $J^0_{\mu 5}$.

Ceci reste standard mais le d\'eveloppement en s\'erie de $\gamma(\alpha_s)$
ne commence qu'au second ordre [KOD] {\it i.e.}
\begin{equation}
\gamma(\alpha_s) = \gamma_2\Biggl({\alpha_s\over 4\pi}\Biggl)^2 + \ldots
\end{equation}
avec
\begin{equation}
\gamma_2 = 16 N_f \,.
\end{equation}

La solution de cette \'evolution est donc
\begin{equation}
\Biggl[\Delta \Sigma-{3\alpha_s\over 2\pi}\Delta g\Biggl]_{Q^2}=
\Biggl[\Delta\Sigma-{3\alpha_s\over 2\pi}~\Delta g\Biggl]_{Q^2_0}
\times \exp \, \left\{ {\gamma_2\over 4\pi\beta_0}\,[\alpha_s(Q^2) -
\alpha_s(Q^2_0)] \right\}
\end{equation}
o\`u on a not\'e comme d'habitude:
$\beta_0 = 11 - {2\over 3}\,N_f\,$.

La quantit\'e
\begin{equation}
\Biggl[\Delta\Sigma -3{\alpha_s(Q^2)\over 2\pi}\Delta g(Q^2)\Biggl]
\times\exp \, \left\{-{\gamma_2\over 4 \pi\beta_0}\,\alpha_s(Q^2)\right\}
\end{equation}
est donc ind\'ependante de $Q^2$. Appelons la $C$.

Dans la limite $Q^2 \to \infty$, comme  $\Delta \Sigma$ est ind\'ependante
de $Q^2$ et que $\alpha_s(Q^2)\to 0$ on obtient:
\begin{equation}
\lim_{Q^2\to\infty}~{3\alpha_s(Q^2)\over 2\pi}~\Delta g(Q^2)=
\Delta\Sigma-C\,.
\end{equation}
Donc  $\Delta g (Q^2)$ doit cro\^\i tre comme
$[\alpha_s(Q^2)]^{-1}$ i.e $\ln (Q^2)$ lorsque $Q^2$ augmente.

\subsection{conclusion}

L'anomalie a donc g\'en\'er\'e une  interaction ponctuelle effective entre
 le photon  virtuel et les gluons et la surprenante petite valeur de $a_0$
peut \^etre comprise comme venant de la compensation entre $\Delta\Sigma$
et la contribution $Q^2$-d\'ependante des gluons via la combinaison
 $(3\alpha_s(Q^2)/2\pi)\,\Delta g(Q^2)$.

Quantitativement si on prend  $a_0 \simeq 0.17$ \`a $Q^2=10$ (GeV/c$)^2$ et
$\alpha_s \simeq 0.24$ on peut consid\'erer que les  quarks portent quelques
 60\% du spin du   proton
 {\it i.e.} choisir $\Delta\Sigma$ = 0.6 et obtenir
\begin{equation}
\Delta g \,[Q^2 = 10~\mbox{\rm (GeV/c)}^2] \simeq 3.8 \,.
\end{equation}
M\^eme si cela peut para\^\i tre trop grand, il ne faut peut-\^etre pas y
attacher
une trop grande signification physique puisque cette valeur d\'epend
crucialement
de $Q^2$. Un exercice pour s'en convaincre et d'\'evoluer ce
$\Delta g (Q^2)$ jusqu'\`a une \'echelle plus proche du r\'egime du Mod\`ele
des
Quarks.
Si on ose descendre jusqu'\`a $Q^2=4\Lambda^2_{\rm QCD}$ o\`u
$\alpha_s \simeq 1$, on trouve
\begin{equation}
\Delta g (4 \Lambda^2_{\rm QCD}) \simeq 0.7 \,,
\end{equation}
ce qui est plus raisonnable.

 M\^eme si on comprend beaucoup
 mieux maintenant la situation, on ne peut \^etre satisfait du tableau
d'ensemble
puisqu'on ne sait toujours pas o\`u se cache le spin du proton. D'autres
exp\'eriences
compl\'ementaires, hors de la diffusion compl\`etement inclusive profonde, sont
n\'ecessaires pour par exemple mesurer la contribution des gluons.
L'\'electroproduction
de saveurs lourdes est une piste possible.

Sur le plan th\'eorique, les r\'esultats exp\'eimentaux ont \'et\'e l'occasion
d'une intense activit\'e. Maintenant que le r\^ole de l'anomalie est
clarifi\'e,
les th\'eoriciens s'attachent \`a comprendre dans des mod\`eles non
perturbatifs
encore bien imparfaits la partition du spin du proton entre ses constituants.
Des estimations dans le mod\`ele de Skirme aux calculs sur r\'eseaux, le
panorama
est riche et divers, et encore en pleine \'evolution.

\section{spin transverse}

L'histoire du spin transverse [AM,CPR], ou comme certains [JJ] pr\'ef\`erent
l'appeler, de
la transversit\'e, a \'et\'e l'occasion, pire encore que celle de
l'h\'elicit\'e rappel\'ee
plus haut, d'\'etonnants contre-sens et oublis. La place pr\'epond\'erante de
la diffusion in\'elastique profonde lepton-nucl\'eon (compl\`etement inclusive)
 et de la description dans le cadre du d\'eveloppement en produits
d'op\'erateurs, dans la maturation de la
description des processus durs en chromodynamique quantique a fait oublier
que d'autres op\'erateurs polaris\'es existaient \`a c\^ot\'e de l'h\'elicit\'e
et que rien n'emp\'echait de parler de partons dans des \'etats de spin
transverse
(transverse signifie toujours par rapport \`a la direction de propagation
du nucl\'eon).

\subsection{definitions}

Il n'est pas inutile de rappeler qu'on peut d\'efinir deux \'etats de spin
 transverse pour une particule de spin $1/2$ \`a partir des \'etats
d'h\'elicit\'e
\ket{+} et \ket{-} par:
\begin{eqnarray}
\ket{+ x} =  \frac {1} {2^{1/2}} ( \ket{+} + e^{i\phi} \ket{-} ) \cr
\ket{- x} =  \frac {1} {2^{1/2}} ( \ket{+} - e^{i\phi} \ket{-} )
\end{eqnarray}
\noindent
o\`u $\phi$ est un  azimuth qui d\'epend  un peu des conventions. Cela implique
que les observables "de spin transverse" joueront le r\^ole d'observables "de
renversement d'h\'elicit\'e ". Or un renversement d'h\'elicit\'e est
notoirement
difficile pour un fermion sans masse et le spin transverse est ainsi un
r\'ev\'elateur
de la brisure de la sym\'etrie chirale dans la chromodynamique r\'ealiste
(celle o\`u
les quarks sont ce qu'ils sont). Un op\'erateur relevant sera donc
n\'ecessairement
impair sous une transformation de chiralit\'e.

Si dans le cas de l'h\'elicit\'e, les densit\'es partoniques \'etaient
d\'efinies \`a
partir des \'el\'ements de matrice du courant axial, comme

\begin{equation}
2MS_\mu [\Delta q_f (x) + \Delta \bar{q}_f (x)]
= \int dy^- e^{ixy^-}\langle P,S|\bar{\psi}_f(0) \gamma_\mu \gamma_5
 \psi_f(y)|P,S \rangle ,
\end{equation}
pour les densit\'es transverses, on a un courant diff\'erent:

\begin{equation}
2MS_x [\Delta q_T (x) - \Delta \bar{q}_T (x)]
= \int dy^- e^{ixy^-}\langle P,S|\bar{\psi}_f(0)(\gamma_0 + \gamma_z) \gamma_x
 \gamma_5 \psi_f(y)|P,S \rangle.
\end{equation}
Le nombre de matrices $\gamma$ explicite le caract\`ere chiral de cette
densit\'e
transverse. Le signe entre la contribution des quarks et celle des antiquarks
marque son caract\`ere impair par conjugaison de charge.

La d\'etermination non perturbative des densit\'es polaris\'ees
transversalement
de quarks et de gluons peut {\it a priori} se faire par calcul sur les
r\'eseaux
comme expliqu\'e par O. P\`ene dans son cours pour la quantit\'e $\Gamma_1
$. Il
suffit de prendre l'op\'erateur ad\'equat indiqu\'e ci-dessus. Bien qu'il
n'existe
actuellement, \`a ma connaissance, aucun r\'esultat, une d\'etermination
 exp\'erimentale am\`enerait sans aucun doute les th\'eoriciens \`a
redoubler leurs efforts dans ce sens.

Notons tout de m\^eme l'existence d'une borne d\'eduite de consid\'erations de
 positivit\'e et reliant densit\'e transverse et densit\'e longitudinale[SO].

\subsection{corrections radiatives}
Les corrections radiatives s'\'etudient de la mani\`ere habituelle et aucune
subtilit\'e ne vient compliquer l'application d'\'equations d'\'evolution de
type Altarelli-Parisi. En particulier et contrairement au cas longitudinal,
aucune anomalie ne vient m\'elanger de fa\c con \'etonnante gluons et partie
singulet des quarks.

\subsection{recherche exp\'erimentale}
Diff\'erents axes de recherche ont \'et\'e propos\'es pour lever le voile de
la derni\`ere quantit\'e (au twist dominant) d\'ecrivant le proton. La plus
prometteuse est sans aucun doute la production de Drell Yan, production d'une
 paire de leptons de grande masse invariante $Q^2$ dans la collision de deux
hadrons polaris\'es transversalement[CPR]. Cet axe sera explor\'e de fa\c con
prioritaire au RHIC de Brookhaven dans sa version de collisionneur
proton-proton
polaris\'es et on peut donc raisonnablement s'attendre aux premiers r\'esultats
sur la contenu en spin transverse des protons vers l'an 2000.

On peut aussi tenter de combiner
polarisation initiale et polarisation finale dans des r\'eactions
lepton-nucl\'eon
 semi-inclusives[HERMES,ELFE] o\`u par exemple un $\Lambda$ serait produit;
on a aussi propos\'e
des quantit\'es construites \`a partir des impulsions des m\'esons de l'\'etat
final (la variable de "handedness" de Nachtmann et Efremov {\it et al}, la
variable de Collins...)[NEC] Mais il faudrait alors
s'assurer que la fragmentation ne dilue pas trop l'information sur le spin
transverse du parton produit ou diffus\'e.

\end{document}